# Hall coefficient in amorphous alloys: critical behavior and quantitative test of quantum corrections due to weak localization and electron-electron interactions

A. Rogachev,[1] H. Ikuta,[2] and U. Mizutani[3]

[1] Department of Physics and Astronomy, University of Utah, Salt Lake City, Utah 84112, USA
[2] Department of Materials Physics, Nagoya University, Nagoya 464-8603, Japan
[3] Nagoya Industrial Science Research Institute, Nagoya, Japan

**ABSTRACT**

The behavior of the Hall coefficient, $R_H$, across the metal-insulator transition in disordered systems is a long-standing unresolved problem. Here, we present the measurements of $R_H$ in a series of $Ti_xSi_{100-x}$ amorphous alloys in temperature range 1.8-300 K. The samples cover a wide range of Ti content reaching the critical concentration, $x_c \approx 9 - 9.5$. Far from the transition, for $x \geq 17$, the Hall coefficient displays the behavior predicted by the perturbation theory, $R_H^{-1}(T) = R_H^{-1}(0) + bT^{1/2}$, which extends up to the temperature 150 K. The temperature dependence gets stronger in alloys with lower $x$; $R_H(0)$ diverges at $x_c$ displaying critical behavior. We used the combined conductivity and Hall coefficient data for alloys with high Ti content to test the theories of quantum corrections to conductivity. We found that the correction due to weak localization is dominated by the electron-phonon scattering with the rate varying with temperature as $\tau_{ep}^{-1} = A_{ep}T^2$. The extracted parameter $A_{ep}$ is in good agreement with the theory that considers the incomplete drag of impurities by lattice vibrations. The spin-orbit scattering time extracted from the weak localization correction was found to be two orders of magnitude larger than the time given by the standard estimate $\tau_{so} \approx \tau(\hbar c/e^2 Z)^4$. This is in accord with observations in several other systems. The theory of the EEI quantum correction was tested using the Hall coefficient and specific heat data for Ti-Si and $(Ag_{0.5}Cu_{0.5})_{100-x}Ge_x$ amorphous alloys, which allowed us to estimate all microscopic parameters needed by the theory. We found that, within the accuracy of our measurements, the EEI theory works exactly for alloys that follow the free electron model [$(Ag_{0.5}Cu_{0.5})_{100-x}Ge_x$ with $x \leq 50$.] The deviation from the theory observed in all Ti-Si alloys and in Ag-Cu-Ge alloys with $x \geq 60$ can be qualitatively explained by weakening of the electron screening in the systems. In a companion paper, we show that in $Ti_{9.5}Si_{90.5}$ alloy, the electronic specific heat coefficient $\gamma$ acquires temperature-dependent critical behavior below 2 K and displays critical behavior.

## I. INTRODUCTION

In the free-electron model, the carrier concentration uniquely defines three experimental physical quantities: single-particle density of states $N(E)$ determined in tunneling experiments; electronic specific heat coefficient $\gamma$; and Hall coefficient $R_H$. This simple picture, however, changes in disordered systems close to and across the metal-insulator transition (MIT). Due to enhanced electron-electron interactions (EEI), $N(E)$ goes to zero at the critical point of the transition while $\gamma$, in the systems studied so far, was found to vary smoothly and non-critically across MIT. These distinct variations have been observed both in doped semiconductors[1,2,3] and amorphous metal-(Si,Ge) alloys [4,5,6,7], two groups of materials often used in the studies of MIT.

The behavior of the Hall coefficient near MIT is not well-understood and is a subject of a long-standing controversy. Even for the non-interacting systems, there exist conflicting predictions for smooth [8,9] and divergent [10,11] variation. For systems with EEI, one anticipates that the temperature dependence of the Hall coefficient obtained in the perturbation theory [12] gets stronger near MIT and leads to divergent behavior. Still, there is an alternative suggestion that $R_H$ follows the noncritical variation of $\gamma$ [13].

Experimentally, the Hall coefficient was found to diverge in Ge:Sb [14], Si:B [15], Si:P [16] and magnetic amorphous alloy $Gd_xSi_{1-x}$ [17] but apparently remains finite in Si:As [13] and $In_2O_{3-x}$ [18]. What determines the difference between two groups remains unclear. The interpretation of experimental results is further complicated in materials containing transition metal elements where spin-orbit scattering [19], electron hybridization [20], diamagnetic currents [21] and magnetism can drastically alter the magnitude and sign of the Hall coefficient. Related problems are also present in organic semiconductors [22,23] and granular materials [24,25,26] where, similar to other disordered systems, the behavior of the Hall coefficient is complex and determined by interplay between coherent electron propagation, hopping transport processes and electron-electron interactions.

In the first part of the paper, we report the measurement of the Hall effect in a series of $Ti_xSi_{100-x}$ ($x=9$ - 32) amorphous alloys spanning MIT. The Hall coefficient in this system was found to be positive. Similar to Si:P and Si:B, it behaves critically at the transition. However, much higher carrier concentration in Ti-Si results in a more pronounced EEI-induced temperature dependence of $R_H$, which in Ti-Si extends up to 150 K, exceeding the corresponding value in doped semiconductors by two-orders of magnitude. This system, therefore, is much better suited for the study of the effects beyond the perturbative quantum corrections; indeed, close to MIT we have observed deviations from perturbative variation. In the second part of the paper, we use the Hall coefficient data to extract the quantum correction due to weak localization from which we, in turn, deduce scattering rates of several relaxation processes in alloys and compare them with theory predictions. In the last part of the paper, we compare the prediction of the Altshuler-Aronov theory of the quantum correction caused by electron-electron interaction [12] with the experimental behavior of the Hall coefficient in Ti-Si and Au-Cu-Ge amorphous alloys. Both systems are well characterized and belong to a very limited group of disordered systems in which the $T$-dependence of the Hall coefficient has been measured [27].The availability of the Hall coefficient data as well as supporting data on specific heat [28,7] allows us to carry out a quantitative test of the theory without any unknown or adjustable parameters.

Let us finally note that in a companion paper [29], we report the measurement of the specific heat in $Ti_{9.5}Si_{90.5}$ alloy down to 0.3 K. We found that the electronic coefficient γ becomes $T$-dependent below 1.5 K and extrapolates to zero at lower temperatures. Ti-Si alloys thus demonstrate the hierarchy of EEI effects: one-electron excitations, as manifested by $R_H$, are affected by EEI at $T \approx 150$ K and below, while the effect of EEI on many-body reorganizations of the systems, as manifested by $\gamma$, are present at much lower ($T \leq 1.5$ K) temperatures.

## II. RESULTS AND DISCUSSION
### a. Critical behavior of the Hall coefficient in $a$-$Ti_xSi_{100-x}$ alloys

The $a$-$Ti_xSi_{100-x}$ samples were deposited on water-cooled glass substrates by DC triode sputtering from composite Ti-Si targets prepared by arc-melting. The typical thickness of the films was about 10 μm. The amorphous structure of the alloys was confirmed by x-ray diffraction; their composition was determined by energy dispersive x-ray analysis with accuracy ±0.4 at.%.

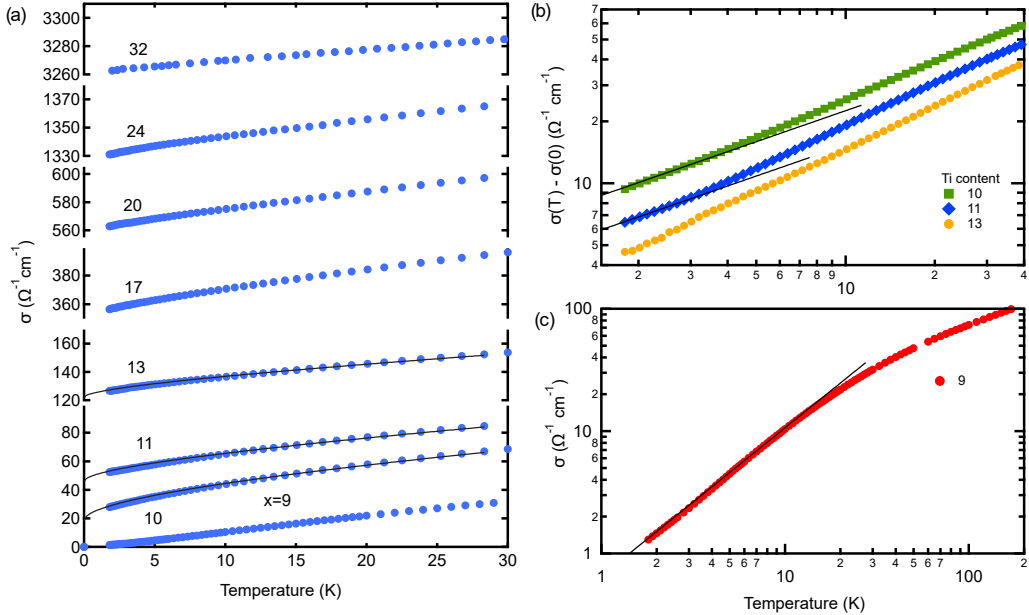

*Fig. 1. (a) Conductivity as a function of temperature for several a-$Ti_xSi_{100-x}$ samples with indicated content of Ti. For alloys with x=10, 11, and 13, the solid lines are fits to Eq. 1 in the text. (b) Temperature-dependent part of the conductivity versus temperature on log-log scale for indicated alloys. Solid lines indicate $T^{1/2}$ dependence. (c) Conductivity versus temperature on log-log scale for alloy with x=9. Solid line represents $T^{1.2}$ dependence.*

Conductivity of the films as a function of temperature was measured using Quantum Design PPMS and is presented in Fig. 1(a). As expected for a strongly disordered system, conductivity of the films increases with temperature. On the metallic side of the transition, this dependence comes from quantum corrections and can be presented as

$$\sigma(T) = \sigma(0) + \sigma_{ee}(T) + \sigma_{wl}(T), \qquad (1)$$

where $\sigma(0)$ is the conductivity at zero temperature, $\sigma_{ee}(T)$ is the correction due to EEI and $\sigma_{wl}(T)$ is due to weak localization. In the studied temperature range, 1.8-300 K, the film thickness is much larger than the dephasing or thermal length in $a$-Ti$_x$Si$_{100-x}$, hence the electron transport has 3d character. In this case, the EEI correction takes a simple form $\sigma_{ee}(T) = \alpha T^{1/2}$. The weak localization correction can be fairly complex; as a starting point of the analysis we take it as $\sigma_{wl}(T) = \beta T$, which ignores the spin-orbit scattering and assumes that the dephasing processes are dominated by electron-phonon scattering [30].

We found that Eq. 1 with square-root and linear terms representing quantum corrections provides good fits to our data. However, the validity of this analysis and extracted fitting parameters ($\sigma(0), \alpha, \beta$) appears questionable. As a test, we used the extracted value of the residual conductivity $\sigma(0)$ to obtain its temperature-dependent part, $\Delta\sigma(T)$, and plotted this quantity versus temperature in Fig. 1(b) on a log-log scale. For many systems, and in particular for amorphous Ag-Cu-Ge alloys [27,28], such plots reveal two branches: one with the slope ½ and the other, at higher temperatures, with the slope 1, confirming chosen dependence of the quantum corrections. This is not the case for $a$-Ti$_x$Si$_{100-x}$ alloys. As shown in Fig.1(b), all we can identify is the presence of a $T^{1/2}$ term at low temperatures in alloys with $x$=10 and 11. Clearly, the linear temperature dependence does not provide a good approximation for the weak localization correction in Ti-Si alloys. We will come back to this question after the discussion of the Hall coefficient.

Despite its shortcomings, the analysis outlined above allows us to identify samples with $x \geq 10$ to be on the metallic side of the MIT. The conductivity of the alloy with $x = 9$ shown in Fig.1(c) displays dependence $\sigma(T) = uT^{1.2}$ phenomenologically expected right at the critical point of the MIT. We, however, cannot exclude the possibility that at lower temperatures the conductivity of this sample switches to an insulating behavior. Therefore, we place the critical concentration at $x_c \approx 9 - 9.5$. This choice is further supported by our previous observation of hopping conductivity in an alloy with $x = 8$ [31].

The Hall voltage was measured by the conventional five-probe AC technique at frequency 13 Hz in the temperature range 1.8-300 K and in magnetic field up to 9 T. To eliminate the magnetoresistance contribution, the sample was rotated by 180 degrees at a fixed temperature and field and half of the difference between the voltages in two orientations was taken to determine the Hall resistivity, $\rho_{xy}$. This procedure was repeated several times to improve accuracy of the measurements. The inset in Fig.2(a) shows the dependence of $\rho_{xy}$ versus field for an alloy with $x$=10. In contrast to the strong non-linear dependence of magnetoresistance (not shown), the Hall resistivity was always found to be linear in magnetic field.

The Hall coefficient $R_H = \rho_{xy}/H$ of all studied Ti-Si films was found to be positive. A positive $R_H$ is fairly common in amorphous alloys containing transition metals [32]. There is, however, no consensus on its physical origin. Some works have claimed that the positive $R_H$ comes from skew scattering induced by spin-orbit interaction [19]. An alternative explanation was put forward by Itoh and Tanaka [20,21], who suggested that in strongly disordered systems one should abandon the $k$-space description and interpret the Hall coefficient in terms of hybridization between orbitals. Particularly, the diamagnetic current is essential for the sign. Itoh and Tanaka also argued that, although $R_H$ reflects the states at the Fermi level, the usual formula $R_H = -1/ne$ is not applicable.

The temperature dependence of the Hall coefficient was measured at fixed magnetic field of 5 or 7 T; it is shown in Fig. 2(a). One can see that the strong variation of $R_H$ appears at low temperatures and $R_H$ increases rapidly with decreasing Ti content. A unique characteristic of the Hall coefficient is that the quantum corrections related to weak localization are exactly zero. The correction due to EEI in the first order of the perturbation theory [12] is related to the conductivity in a simple way

$$\delta R_H/R_H = -\delta R_H^{-1}/R_H^{-1} = -2\delta\sigma_{xx}/\sigma_{xx}. \qquad (2)$$

This means that similar to $\sigma_{ee}$, the Hall coefficient is expected to follow simple $T^{1/2}$ dependence.

In Fig. 2(b,c) we plotted the inverse of the Hall coefficient versus $T^{1/2}$. This figure presents the central result of our study. Far from the metal-insulator transition, for an alloy with x=32, temperature dependence of $R_H$ is not resolved. Alloys with higher disorder and lower carrier density, which are still located quite far from the MIT (x=24,20,17) display temperature dependence expected from the perturbation theory, $R_H^{-1}(T) = $ $R_H^{-1}(0) + bT^{1/2}$, covering a very broad temperature range 1.8-150 K. Alloys with $x \leq 13$, as we argued in Ref. [7], are already in the scaling regime where the temperature dependence at low temperatures is determined by the specific form of the scaling function characterizing the transition. At higher temperatures, perturbative behavior is still expected. Guided by this picture, we fitted the data for alloys with $x \leq 13$ at high temperatures with $T^{1/2}$ dependence and extended these fits (shown in the figure as solid lines) to lower temperatures. From these plots, one can see that in the scaling regime, the inverse of the Hall coefficient deviates from the perturbation theory prediction and shows stronger temperature dependence. This behavior becomes even more pronounced in the insulating sample with $x = 9$ (Fig. 2a,c), where $R_H$ displays strong divergent temperature dependence and becomes undetectable at temperatures below 10K.

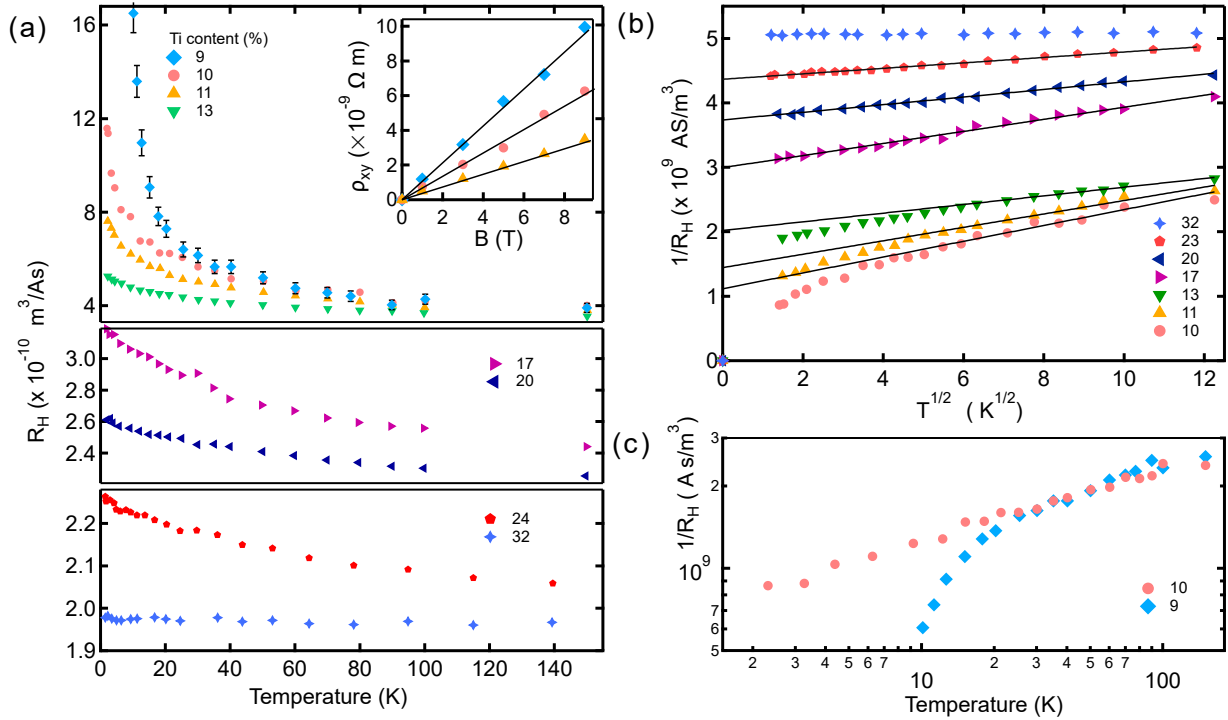

Fig.2 (a) Hall coefficient as function of temperature for the studied series of Ti-Si amorphous alloys. Inset shows the dependence of the Hall resistivity for an alloy with x=10 at several temperatures. (b) The inverse of the Hall coefficient versus $T^{1/2}$. Solid lines show fits to $R_H^{-1}(T) = R_H^{-1}(0) + bT^{1/2}$ dependence. (c) The inverse of the Hall coefficient versus temperature on log-log scale for alloys with $x = 9$ and 10.

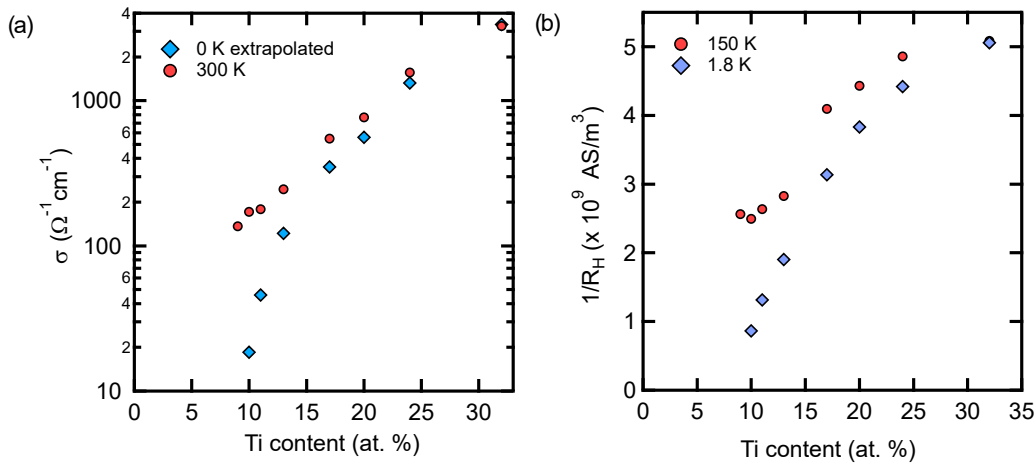

Fig. 3 Conductivity (a) and inverse Hall coefficient (b) as a function of the Ti content at indicated temperatures.

The variation of the conductivity and inverse Hall coefficient at low and high temperatures versus Ti content in the alloys is shown in Fig. 3. Both quantities display critical behavior. We can say with sufficient confidence that the critical behavior of $R_H$ is caused exclusively by EEI. The presence of the critical point by itself does not change the microscopic processes in a system, so it is unlikely that localization processes absent in $R_H$ for alloys with $x \geq 17$, come back for alloys with lower concentration of Ti. We have refrained ourselves from numerical evaluation of the critical exponents because our data do not go to low enough temperatures. In doped semiconductors, Si:P [16] and Si:B[15], the Hall correlation length exponent was found to be $\nu_H \approx 1/2$. In our view, this is not a reliable result. It is based on the extrapolation of the data to zero temperature with $T^{1/2}$ dependence. However, the large scattering of the data points in original papers does not make the presence of this dependence obvious in the first place. Moreover, our results show that deviations from $T^{1/2}$ at low temperatures close to MIT are quite likely.

**b. Quantitative test of the weak localization correction. Inelastic and spin-orbit scattering times.**

Conductivity of a strongly disordered metal is determined by two quantum corrections, $\sigma_{wl}$ and $\sigma_{ee}$. The theoretical description of both corrections is well-established; however, the final formulas depend on several, typically-unknown, microscopic parameters. In practice, this often means that 4-6 adjusting coefficients are used to fit a fairly featureless temperature dependence of conductivity [33,34]. This procedure raises concerns as to whether the extracted parameters truly represent a system. The situation can be further complicated by the presence of superconducting fluctuations. In fact, an accurate distinction between superconducting fluctuations and quantum corrections related to normal electrons has become crucial for the recently discovered pair-breaking quantum phase transitions [35]. These transitions are induced by a magnetic field and take place only in a superconducting subsystem of a nanowire or film in the presence of the dominant non-critical conductivity of normal electrons. Ideally, one would like to compute this conductivity from known or estimated microscopic parameters so it could be then subtracted to reveal the critical contribution of superconducting fluctuations.

Unlike most of the systems studied so far, availability of the specific heat, magnetic susceptibility and most importantly Hall coefficient makes Ti-Si an excellent candidate to address these problems. EEI correction is analyzed in the next section. Here, we capitalize on our observation that the Hall coefficient for alloys with $x = 17, 20, 24$ accurately follows the $T^{1/2}$ dependence and use Eq. 2 to extract the weak localization correction as following, $\sigma_{wl}(T) = \sigma(T) - \sigma(0) - \sigma_{ee}(T) = \sigma(T) - \sigma(0) - bT^{1/2}\sigma(0)/2R_H^{-1}(0)$. Here, $bT^{1/2} = \delta R_H^{-1}(T)$ is the experimental temperature-dependent part of the Hall coefficient. The obtained conductivity is shown in Fig. 4. We found that small variations of the extrapolated $\sigma(0)$ and $R_H^{-1}(0)$ do not change the shape of the curve $\sigma_{wl}(T)$ but simply shift it up or down.

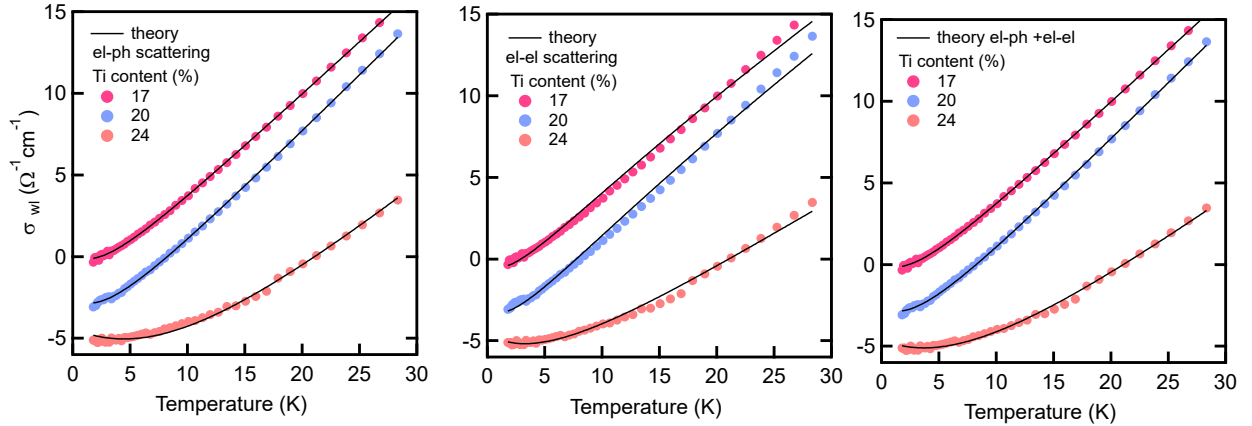

*Fig.4. The weak localization correction to conductivity for three Ti-Si amorphous alloys. The solid lines are the theoretical fits.*

Table 1 presents several measured or estimated parameters of Ti-Si alloys needed for the quantitative analysis of the data. The density of states, $g(0)$, was computed from measured specific heat and mass density of the alloys, $\rho_M$. The diffusion coefficient was obtained from equation $\sigma = e^2 D g(0)$ using conductivity at 300 K. This conductivity was also used to estimate the Fermi vector, $k_F$, from the equation $\sigma = e^2 \ell k_F^2 / 3\pi^2 \hbar$. The mean free path $\ell$ was taken to be equal to nearest-neighbor interatomic distance of 0.26 nm determined from X-ray diffraction data. Using standard relations, $D = v_F \ell/3$, $v_F = \hbar k_F/m^*$, $\ell = v_F \tau$, and $k_F = (3\pi^2 n)^{1/3}$, the elastic scattering time $\tau$, effective mass $m^*$ and carrier density $n$ were estimated.

**Table 1**

| Ti (at.%) | $\sigma$ (300K) ($\Omega^{-1}cm^{-1}$) | $\sigma$ (0 K) ($\Omega^{-1}cm^{-1}$) | $g(0)$ ($J^{-1}m^{-3}$) | $D$ ($cm^2 s^{-1}$) | $k_F$ ($m^{-1}$) | $k_F \ell$ | $m^*/m_e$ | $\tau$ (s) | $n$ ($m^{-3}$) |
|---|---|---|---|---|---|---|---|---|---|
| 17 | 350 | 550 | $1.0 \times 10^{47}$ | 0.215 | $5.3 \times 10^9$ | 1.1 | 2.3 | $10^{-15}$ | $4.4 \times 10^{27}$ |
| 20 | 560 | 770 | $1.2 \times 10^{47}$ | 0.255 | $6.2 \times 10^9$ | 1.3 | 2.3 | $8.9 \times 10^{-16}$ | $7.3 \times 10^{27}$ |
| 24 | 1330 | 1560 | $1.75 \times 10^{47}$ | 0.28 | $7.9 \times 10^9$ | 1.7 | 2.6 | $8 \times 10^{-16}$ | $2.1 \times 10^{28}$ |

*Several physical quantities characterizing Ti-Si alloys with indicated Ti content.*

Theoretically the weak localization correction depends on inelastic scattering time $\tau_{in}$, spin-scattering time $\tau_s$ and spin-orbit scattering time $\tau_{so}^{wl}$. (Here we introduce the super-index to distinguish it from usual $\tau_{so}$). Measurements of magnetic susceptibility have indicated that Ti-Si amorphous alloys with $x = 17, 20, 24$ are non-magnetic; Ti does not carry a magnetic moment in this system [36]. We therefore can neglect spin scattering and the theoretical formula for $\sigma_{wl}$ becomes [33]

$$\sigma_{wl}(T) = \frac{e^2}{2\pi\hbar} \frac{1}{\sqrt{D}} \left[ 3 \left( \frac{1}{4\tau_{in}(T)} + \frac{1}{3\tau_{so}^{wl}} \right)^{1/2} - \left( \frac{1}{4\tau_{in}(T)} \right)^{1/2} \right], \qquad (3)$$

To make a comparison with the experiment, a constant was added to the equation to compensate for up and down shifts of the curves as explained above. Comparison with the theories presented below suggests that, in the studied temperature range, there are comparable contributions to $\tau_{in}$ from electron-phonon scattering, which depends on temperature as $1/\tau_{ep} = A_{ep} T^2$, and from electron-electron scattering, for which $1/\tau_{ee} = B_{ee} T^{3/2}$. We therefore carried out three independent fitting procedures approximating inelastic scattering as (i) $1/\tau_{in} = 1/\tau_{ep}$, Fig. 4a, (ii) $1/\tau_{in} = 1/\tau_{ee}$, Fig. 4b, and (iii) $1/\tau_{in} = 1/\tau_{ep} + 1/\tau_{ee}$, Fig. 4c, taking $A_{ep}, B_{ee}, \tau_{so}^{wl}$ as adjustable parameters. One can see from the figure that considering the electron-phonon scattering alone already gives good approximation to the data; fits with electron-electron scattering are less satisfactory, and, as expected, taking both processes into account gives the best fits to the data. The parameters obtained in the latter (iii) case are presented in Table 2 and compared with the theories.

**Table 2**

| Ti | $1/\tau_{ep}$(exp) (1/s) | $1/\tau_{ep}^d$ (theo) (1/s) | $1/\tau_{ep}^{qs}$ (theo) (1/s) | $1/\tau_{ee}$ (exp) (1/s) | $1/\tau_{ee}$ (theo) (1/s) | $\tau_{so}^{wl}$ (exp) (s) | $\tau_{so}^{ss}$ (theo) (s) |
|---|---|---|---|---|---|---|---|
| 17 | $5.8 \times 10^7 T^2$ | $6 \times 10^6 T^2$ | $2.2 \times 10^7 T^2$ | $3.0 \times 10^7 T^{1.5}$ | $2.5 \times 10^8 T^{1.5}$ | $1.3 \times 10^{-9}$ | $6.6 \times 10^{-12}$ |
| 20 | $7.7 \times 10^7 T^2$ | $9.5 \times 10^6 T^2$ | $3.5 \times 10^7 T^2$ | $6.5 \times 10^7 T^{1.5}$ | $1.7 \times 10^8 T^{1.5}$ | $7.1 \times 10^{-10}$ | $5.3 \times 10^{-12}$ |
| 24 | $1.1 \times 10^8 T^2$ | $2.6 \times 10^7 T^2$ | $6.0 \times 10^7 T^2$ | $1.1 \times 10^8 T^{1.5}$ | $9.7 \times 10^7 T^{1.5}$ | $1.2 \times 10^{-10}$ | $4.4 \times 10^{-12}$ |

*Experimental and theoretical parameters characterizing quantum correction due to weak localization. Temperature is in Kelvins. "(exp)" and "(theo)" comments indicate that a parameter was extracted from the fits or computed theoretically.*

Theory predicts that the enhancement of the electron-phonon scattering in disordered metals comes from two processes. In the first process, the impurity atoms are dragged together with the lattice by phonons and thus change electron-phonon interaction. Bergmann and Tokayama analyzed this process and found that at low temperatures its rate varies as [37]

$$\frac{1}{\tau_{ep}^d} = 2.47 \frac{n}{\rho_M m} \frac{1}{v_l^3} \frac{1}{\ell} (k_B T)^2 \qquad (4)$$

In our analysis, we use the values for the longitudinal, $v_l = 7.96 \, km/s$, and transverse, $v_t = 3.85 \, km/s$, sound velocities in amorphous silicon computed in Ref. [38]. We use super-index "d" to indicate that this contribution comes from "dragged" impurities. More recently, Sergeev and Mitin [39] realized that there is an additional scattering process caused by an incomplete drag. The effect is strong in materials composed of elements with distinct masses (which is the case of our Ti-Si alloys) and is characterized by a coefficient $k \equiv 1 - \ell/\mathcal{L}$. Here, $\mathcal{L}$ is an electron mean free path caused by the displacement of the atoms due to incomplete phonon drag; we use notation from Ref. [40]. Presence of this strong second contribution can make electron-phonon scattering a dominant inelastic process down to very low temperatures. The temperature range of our experiments falls into the diffusive limit of the theory, $q_t \ell \approx k_B T \ell / \hbar v_t \ll 1$. (Here $q_t$ is the wave number of transverse phonons).

In this limit, the electron-phonon scattering time is dominated by $T^2$ term and given by the following equation [39]

$$\frac{1}{\tau_{ep}^{qs}} = \frac{3\pi^2(k_B T)^2 \beta_l}{(\hbar k_F v_t)(\hbar k_F \ell)} k(1-k). \tag{5}$$

Here $\beta_l = (2E_F/3)^2 g(0)/2\rho_M v_t^2$. Following the terminology of Ref. [40] we refer to this contribution as quasistatic and use super-index "$qs$" to indicate this.

The rate of the electron-phonon scattering computed using Eqs. 4 and 5 is shown in Table 2; the incomplete drag coefficient was chosen to be $k = 0.5$. Let us first notice that, in agreement with the reasoning of Sergeev and Mitin, the contribution to $1/\tau_{ep}$ caused by drag is much weaker than the quasistatic contribution. Also, in agreement with the theory prediction, the experimental rate of the scattering increases with Ti content. The numerical discrepancy between the experimental value of the electron-phonon scattering rate and its theoretical prediction, $1/\tau_{ep}^{theo} = 1/\tau_{ep}^{d} + 1/\tau_{ep}^{qs}$, is within a factor of 2 and can be explained by the uncertainties in the parameters of the alloy. In particular, there is a study [41] in which the speed of sound in amorphous silicon prepared by sputtering (the method we used) was found to be 60% of that in crystalline silicon. These reduced values would give an almost exact match to our experiments. Hence, we can conclude that current theories give an accurate quantitative prediction of electron-phonon scattering rate in Ti-Si amorphous alloys. This is in accord with the analysis of $1/\tau_{ep}$ in metallic glass CuZrAl reported in Ref. [40]. We also would like to comment that electron-phonon scattering rate extracted from the weak localization correction appears to be a reliable parameter for the electron thermalization [42].

The second contribution to the inelastic scattering comes from electron-electron scattering. For three-dimensional disordered metal the rate of this process is the same as the dephasing rate and is given by the equation (Eq. 4.4 in Ref.12 )

$$\frac{1}{\tau_{ee}} = C \times 0.0056 \times \frac{1}{\hbar g(0)} \left(\frac{k_B T}{\hbar D}\right)^{3/2}, \tag{6}$$

where $C$ is a constant of order 1, related to the screening in a material. Theoretical values of $1/\tau_{ee}$ for three Ti-Si amorphous alloys are given in Table 2. The theory matches the experimental $1/\tau_{ee}$ in the alloy with $x = 24$. Still, this might be a coincidence since the theory and experiment have opposite dependencies on Ti content. Overall, we can conclude that the theory gives a prediction that agrees with the experiment only by an order of magnitude; discrepancies in alloys with $x = 17, 20$ may possibly relate to their proximity to the critical point of MIT.

Let us finally discuss the third parameter extracted from the fits, the spin-orbit scattering time, $\tau_{so}^{wl}$. As shown in the table, it differs by two orders of magnitude from a standard estimate for the spin-orbit scattering time, $\tau_{so}^{ss} \approx \tau(\hbar c/e^2 Z_{eff})^4$, where the effective atomic number was estimated as $Z_{eff} = xZ_{Ti} + (1-x)Z_{Si}$. The same order discrepancy was also found in amorphous MoGe [43] and CuTi [30] alloys and clearly is systematic. In Ref. [30], the authors argued that two scattering times are different in nature; $\tau_{so}^{wl}$ represents spin-flip relaxation time, while $\tau_{so}^{ss}$ is the spin-orbit relaxation time which "arises in the context of the skew scattering contribution to the Hall effect". We found that $\tau_{so}^{ss}$ is a very useful parameter that gives quantitatively correct values of magnetic-field-induced spin pair-breaking in superconducting Nb and MoGe nanowires [43] and explains anomalous dependence of the pair-breaking strength of magnetic Gd atoms on thickness of superconducting MoGe films [44]. The authors of Ref. [30] found strong, approximately $Z^{-12}$ dependence of $\tau_{so}^{wl}$ induced by changing relative content of Ti and Cu in CuTi alloys. They also presented the theoretical arguments that $\tau_{so}^{wl}$ is the second order effect in interaction and should vary as $\sim Z^{-8}$. In our Ti-Si alloys, the extracted values of $\tau_{so}^{wl}$ exhibit similar, an order of magnitude growth between $x=24$ and 17. We, however, would like to restrain ourselves from making any conclusions from this behavior. Our system is too complex for a quantitative study of the spin-orbit scattering. In addition to the changes of the electronic parameters, alloys with x=24-17 undergo the change in the local atomic structure from one locally resembling Ti disilicides to one with Ti atoms distributed in the matrix of amorphous Si. To conclude, the difference between $\tau_{so}^{wl}$ and $\tau_{so}^{ss}$ is systematic and remains, in our view, not understood. It certainly requires further investigation.

## c. Quantitative test of the correction due to electron-electron interactions

In a three-dimensional disordered metal, quantum correction to conductivity due to electron-electron interaction (EEI) is given by the formula (Eq. 5.6 in Ref. [12])

$$\sigma_{ee}(T) = 0.915 \frac{e^2}{4\pi^2 \hbar}\left(\frac{4}{3} + \frac{3F_\sigma}{2}\right)\left(\frac{k_B T}{\hbar D}\right)^{1/2} = \alpha_{ee} T^{1/2}, \quad (7)$$

where quantity $F_\sigma$ can be expressed as $F_\sigma = (32/3F)\left(1 + 3F/4 - (1 + F/2)^{3/2}\right)$ with $F$ being the exact static scattering amplitude of a particle and a hole with total spin $j = 1$. In the free electron model, quantity $F$ is related to the Fermi wave vector and the Thomas-Fermi screening wave vector, $k_0^2 = e^2 g(0)/\varepsilon_0$, as $F = (1/x)\ln(1 + x)$, where $x = (2k_F/k_0)^2$.

From the above equations, the EEI correction can be computed exactly from the conductivity and density of states, provided the system follows the free electron model. This is certainly not the case in Ti-Si amorphous alloys or, to the best of our knowledge, in any other system used for analysis of the EEI correction. Typically, the quantity $F_\sigma$ is taken as an unknown parameter extracted from experiment. Fortunately, there is an exceptional amorphous alloy system, $(Ag_{0.5}Cu_{0.5})_{100-x}Ge_x$, studied extensively in the past by one of us (UM), that behaves as a free electron metal for a wide range of Ge content. Availability of the Hall coefficient data, which reveals theoretically expected $T^{1/2}$ temperature dependence, makes this system uniquely suited for the quantitative test of the EEI correction. That is what we will do first in this section; then, we will come back to the case of TiSi alloys.

$(Ag_{0.5}Cu_{0.5})_{100-x}Ge_x$ amorphous alloys can be formed by liquid quenching and/or sputtering with any amount of germanium content, $0 \leq x \leq 100$ [28, 45]. Measurements of the specific heat and Hall coefficient indicated that, in the range $x \lesssim 40\ at.\%$, the two quantities agree with each other within the free electron approximation with $m^* = m_e$; moreover, the extracted carrier concentration was found to grow with $x$ according to the simple rule that each Ag and Cu atom donates one free electron and each Ge atom four . At higher Ge content, this tendency breaks down because Ge atoms start to form covalent bonds. The alloy $Ag_{0.5}Cu_{0.5}$ has a fairly long mean free path of 6.3 nm; this quickly decreases with Ge content. For an alloy with $x \approx 30$, the mean free path reaches the minimum possible value, which is equal to the interatomic distance of approximately 0.27 nm. We take this value as an estimate of the mean free path in the alloys of our interest, $30 \leq x \leq 70$, where the quantum EEI correction was resolved in the conductivity and Hall measurements.

In Table 3 we present several parameters characterizing these alloys. To obtain the density of states, $g(0)$, we used experimentally measured specific heat for alloys with $x$=20,40 and 60, and linear extrapolation between these values for alloys with $x$=30, 50 and 70. To obtain the diffusion coefficient, Fermi wave vector and effective mass we used the same equations as in the previous section. The Thomas-Fermi screening wave vector was computed as $k_0^2 = e^2 g(0)/\varepsilon_0$.

**Table 3**

| Ge | $\sigma(300K)$ $1/\Omega\ cm)$ | $\frac{\sigma(300K)}{\sigma(2K)}$ | $g(0)$ $(1/Jm^3)$ | $D$ $(cm^2/s)$ | $k_F$ $(1/m)$ | $k_0$ $(1/m)$ | $m^*/m_e$ |
|----|-------|------|---------|-----|---------------------|---------------------|------|
| 30 | 49500 | 1.05 | 1.2e+47 | 1.6 | $1.5 \times 10^{10}$ | $1.9 \times 10^{10}$ | 0.96 |
| 40 | 37600 | 1.07 | 1.22e+47 | 1.2 | $1.3 \times 10^{10}$ | $1.9 \times 10^{10}$ | 1.1 |
| 50 | 26400 | 1.1 | 8.7e+46 | 1.1 | $1.1 \times 10^{10}$ | $1.6 \times 10^{10}$ | 0.99 |
| 60 | 20000 | 1.14 | 6.2e+46 | 1.25 | $9.5 \times 10^9$ | $1.4 \times 10^{10}$ | 0.78 |
| 70 | 6500 | 1.3 | 3.4e+46 | 0.71 | $5.3 \times 10^9$ | $1.0 \times 10^{10}$ | 0.77 |

*Several physical quantities characterizing $(Ag_{0.5}Cu_{0.5})_{100-x}Ge_x$ amorphous alloys with indicated Ge content.*

The estimates presented above provide us all physical quantities needed for computation of the coefficient $\alpha_{ee}$ in the EEI quantum correction for conductivity, $\sigma_{ee} = \alpha_{ee} T^{1/2}$. The results, alongside with intermediate parameters, $F$ and $F_\sigma$, are presented in Table 4. We estimate the standard deviation of the theoretical value of $\alpha_{ee}$ to be $\pm 20\%$. The experimental values of the same coefficient were extracted from two sets of measurements done on two independent series of samples. Column labeled as "exp1" shows $\alpha_{ee}$ extracted from the temperature dependence of conductivity reported in Ref. [45]; column "exp2" utilized the temperature dependence of the Hall coefficient of Ref. [27]. (Let us note that in Ref. [27] the conductivity was also measured and was found to match $R_H$ as predicted by Eq. 2). Comparison of $\alpha_{ee}$ obtained from two sets of data gives the standard deviation for this quantity at about $\pm 30\%$.

**Table 4**

| Ge | $F$ | $F_\sigma$ | $\alpha_{ee}$ (theo) $(1/\Omega \text{ cm K}^{1/2})$ | $\alpha_{ee}$ (exp1) $(1/\Omega \text{ cm K}^{1/2})$ | $\alpha_{ee}$ (exp2) $(1/\Omega \text{ cm K}^{1/2})$ |
|---|---|---|---|---|---|
| 30 | 0.5 | -0.48 | 1.0 | 0.91 | |
| 40 | 0.56 | -0.54 | 1.0 | 1.2 | |
| 50 | 0.58 | -0.55 | 0.98 | 1.05 | 1.15 |
| 60 | 0.55 | -0.53 | 1.0 | 2.5 | 3.9 |
| 70 | 0.67 | -0.64 | 0.92 | 5.6 | 6.8 |

*Theoretical and experimental values of the parameter $\alpha_{ee}$ in the temperature dependence of the EEI quantum correction for a series of $(Ag_{0.5}Cu_{0.5})_{100-x}Ge_x$ amorphous alloys with indicated Ge content. The theoretical values of the parameters $F$ and $F_\sigma$ are also shown.*

The first three lines in the Table 4 present the central results of this section. A close agreement between the experimental and theoretical values of $\alpha_{ee}$ indicates that the theory works exactly for a material that satisfies the needed conditions, most notably compliance with the free electron model (see last column in Table 3). This, in our view, is a spectacular outcome.

The data in Table 4 also shows that for alloys with high concentration of Ge, the experiment deviates quite strongly from the theory prediction. We have found disagreement of the same order in Ti-Si alloys. Table 5 presents the experimental coefficients, $\alpha_{ee}^{exp}$, obtained for the Hall coefficient and theoretical coefficients, $\alpha_{ee}^{theo}$, computed using Eq. 7 and parameters of Ti-Si alloys summarized in Table 3.

**Table 5**

| Ti | $F$ | $F_\sigma$ | $k_0$ $(1/m)$ | $\alpha_{ee}^{theo}$ $(1/\Omega \text{ cm K}^{1/2})$ | $\alpha_{ee}^{exp}$ $(1/\Omega \text{ cm K}^{1/2})$ |
|---|---|---|---|---|---|
| 17 | 0.85 | -0.79 | $1.7 \times 10^{10}$ | 0.64 | 2.5 |
| 20 | 0.83 | -0.78 | $1.85 \times 10^{10}$ | 0.7 | 5.2 |
| 24 | 0.81 | -0.76 | $2.25 \times 10^{10}$ | 0.72 | 4.24 |

*Theoretical and experimental values of the parameter $\alpha_{ee}$ in the temperature dependence of the EEI quantum correction for a series of $Ti_xSi_{100-x}$ amorphous alloys with indicated Ti content. The theoretical values of the parameters $F$, $F_\sigma$, and $k_0$ are also shown.*

Disagreement with the theory most certainly relates to the fact that the free electron model does not work for all Ti-Si alloys and in Ag-Cu-Ge alloys with germanium content above 50%. One can try to loosen this criterium and, as is often done for amorphous alloys and liquid metals, assume that the Fermi surface remains spherical but $E(k)$ relation changes to be characterized at the Fermi surface by effective mass $m^*$. This approximation apparently does not help to reconcile the discrepancy with experiment, since the effective mass in Ag-Cu-Ge is smaller than one and in Ti-Si is larger than one, while the experimental $\alpha_{ee}$ is larger than the theoretical $\alpha_{ee}$ in both systems.

Likely, the computation of parameter $F$ based on the free electron model is simply not adequate for the alloys with high metalloid content. Indeed, the estimates give the Thomas-Fermi screening length in the range $\lambda_{TF} = 1/k_0 \approx 0.5 - 1$ Å, while the formation of the covalent Ge-Ge or Si-Si bonds modify systems on the larger scale ~2.5 Å, strongly altering local screening. Moreover, since parameter $F$ represents the exact static scattering amplitude of a particle and a hole with total spin $j = 1$, it must be affected by the spin-orbit interaction. Nevertheless, we find the positive deviation from the theory observed in both systems to be physically meaningful. Within the present theory, this tendency can be formally associated with the growth of $\lambda_{TF}$ (decrease of $k_0$) beyond the free electron approximation. So, it indeed reflects the worsening of the electron screening expected in alloys with high content of Ge or Si.

## III. SUMMARY

The Hall coefficient in Ti-Si amorphous alloys has positive sign and displays critical, divergent behavior at the metal-insulator transition. This divergence comes from its temperature dependence and is, in our view, associated exclusively with the enhancement of the electron-electron interactions. Close to transition on the metallic side, the inverse of the Hall coefficient deviates from the behavior predicted by the perturbation theory $R_H^{-1}(T) = R_H^{-1}(0) + bT^{1/2}$, and displays more pronounced temperature variation. Far from the critical point, the alloys show the $T^{1/2}$ perturbative behavior from 150 K down to 1.8 K, the lowest temperature of experiment. Availability of the Hall coefficient data allowed us to independently analyze the quantum corrections due to weak localization and electron-electron interaction. The weak localization correction is dominated by the quasistatic electron-phonon scattering; the rate of this process is quantitatively consistent with the theory of Sergeev and Mitin. For analysis of the EEI correction, we used, in addition to presented data for Hall coefficient in Ti-Si alloys, the data for $(Ag_{0.5}Cu_{0.5})_{100-x}Ge_x$ alloys borrowed from literature. We found that, within the accuracy of our measurements, the theory of the perturbative EEI correction works exactly for alloys that follow the free electron model (Ag-Cu-Ge with $x \leq 50$). Deviation from the theory observed in all Ti-Si alloys and in Ag-Cu-Ge alloys with $x > 60$ can be qualitatively explained by weakening of the screening in these systems.

## Acknowledgements

AR gratefully acknowledges the support by NSF Grant DMR1904221.